# Representations of Sound in Deep Learning of Audio Features from Music

**Sergey Shuvaev, Hamza Giaffar, and Alexei A. Koulakov**

*Cold Spring Harbor Laboratory, Cold Spring Harbor, NY*

## Abstract

The work of a single musician, group or composer can vary widely in terms of musical style. Indeed, different stylistic elements, from performance medium and rhythm to harmony and texture, are typically exploited and developed across an artist's lifetime. Yet, there is often a discernable character to the work of, for instance, individual composers at the perceptual level – an experienced listener can often pick up on subtle clues in the music to identify the composer or performer. Here we suggest that a convolutional network may learn these subtle clues or features given an appropriate representation of the music. In this paper, we apply a deep convolutional neural network to a large audio dataset and empirically evaluate its performance on audio classification tasks. Our trained network demonstrates accurate performance on such classification tasks when presented with ~ 5 s examples of music obtained by simple transformations of the raw audio waveform. A particularly interesting example is the spectral representation of music obtained by application of a logarithmically spaced filter bank, mirroring the early stages of auditory signal transduction in mammals. The most successful representation of music to facilitate discrimination was obtained via a random matrix transform (RMT). Networks based on logarithmic filter banks and RMT were able to correctly guess the one composer out of 31 possibilities in 68% and 84% of cases respectively.

## 1 Introduction

Many auditory stimuli, including those drawn from speech, music and nature, are complex and high dimensional. Learning features from such signals, which often have structure at many timescales, has proven to be challenging. A number of sparse coding models have achieved impressive success in learning relatively shallow features, [1, 2] however there is significantly less work in the area of complex audio feature learning.

In recent years, deep convolutional neural networks have been used with great success in a wide range of contexts. It is perhaps in the area of computer vision that the most striking results have been achieved, particularly in tasks such as object detection and image recognition/captioning [3, 4]. The development and use of deep recurrent and convolutional neural networks (CNNs) in image processing has been inspired by our understanding of the neurobiology of visual cortex and the statistics of visual scenes. Similar success with deep learning approaches is lagging in the domain of audio signals, however CNNs are showing signs that they may also be well suited to the problem of learning musical features, given an appropriate representation of sound.

In this paper, we apply a deep convolutional neural network to a large custom audio dataset representing different signal transformations and empirically evaluate its performance on a large-scale audio classification task. We compare the representation of music obtained via a logarithmic filter bank (log spectrogram) with its linear analogue as well as a representation obtained via a random transform detailed below.

## 2  Methods

### 2.1 Description of custom audio dataset

The dataset used in this paper is composed of 2D representations of music obtained from time varying waveforms. Audio files (.mp3) corresponding to 31 composers of classical music spanning nearly 350 years and a variety of sub-genres were downloaded from youtube.com (~ 2 hours per composer) in compliance with fair use policies.

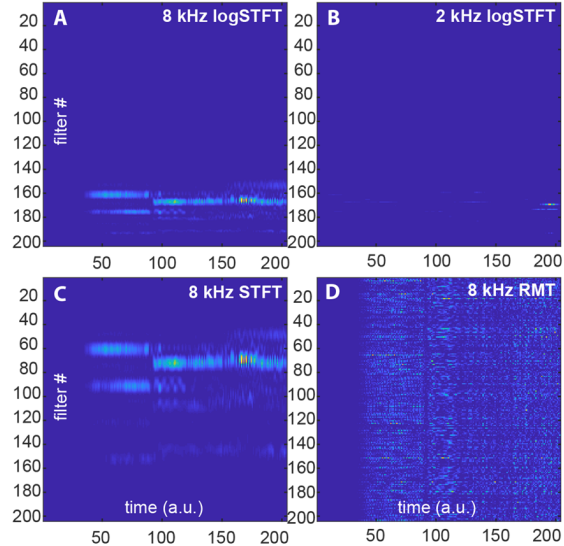

Figure 1 – Representations of music obtained via (A) logSTFT (8kHz dataset), (B) log STFT (2kHz dataset) (C) STFT and (D) RMT both (8kHz dataset)

Audio preprocessing is crucial to this approach. If required, sound files are first reduced to a monophonic signal by averaging over channels of the stereo recording. These audio signals are then down sampled to a sampling rate, Fs = 8000 Hz (~5.2s of music). We also generate a dataset at 2kHz for comparison (~ 20.8s of music per spectrogram). In these analyses, we applied a flat window function to the audio signal and for 1) and 2) we discard phase information and take the absolute value of the square of the amplitude of the transformed signal. The central frequencies of these filters are distributed between $f_{min}$ = 16.35 ($C_0$ double pedal C) and $f_{max}$ = 5587.65 ($F_8$). For transform 1), the filters therefore roughly correspond to the notes of the full chromatic scale together with half tones between each adjacent note. Inputs to the network are 204 x 204 pixels.

The full list of composers: Part, Bach, Beethoven, Chopin, Cage, Liszt, Mozart, Ravel, Schumann, Shostakovich, Clementi, Scarlatti, Haydn, Rachmaninov, Bartok, Brahms, Prokofiev, Schnittke, Schubert, Britten, Glass, Boulez, Gershwin, Vivaldi, Schoenberg, Telemann, Stravinsky, Handel, Martinu, Kodaly, and Nazaykinskaya.

### 2.2 Representation of Sound

2D input to the CNN is generated from raw audio waveforms via three different transformations.

- 1) Logarithmic Short-Time Fourier Transform (logSTFT) - STFT with the filters distributed uniformly on the logarithmic frequency axis.
- 2) STFT - closely related to 1), but with filters that are equally spaced on the linear frequency axis.
- 3) Random Matrix Transform (RMT) – related to 1) and 2), but with a random matrix R (values drawn iid from a normal distribution) in the place of the discrete Fourier transform (DFT) matrix.

The auditory system performs a spectroscopic analysis of sensory signals. At the periphery, the cochlea transduces sound pressure signals and encodes their statistics in neural spike trains. In effecting this transformation, the cochlea can be thought of as a bank of filters, with filter centers distributed equally in the logarithmic frequency scale. Sound, as it is encoded in the activity of early sensory neurons, is therefore well represented in a spectrogram with a logarithmic frequency axis. These spectrograms were generated by applying the discrete short-time Fourier transform (STFT) to samples of the time signal and then taking the square of the magnitude of the result.

It has been suggested that information from auditory signals is compressed into summary statistics, which encode stimulus information efficiently and flexibly. These summary statistics may be computed over a succession of short time windows in order to build a lower dimensional and therefore more efficiently encoded time varying representation of a heterogeneous signal [6, 7]. The STFT effectively computes the frequency content of local sections or frames of a time varying signal and as such the generated spectrograms are reasonable approximations to the output of a bank of cochlear filters. We choose to adopt filters following the frequency tuning of the diatonic scale with the range of a full piano, however an alternative representation of the filtered output of the human cochlea could be achieved by adopting the mel scale. These images will serve as the input to the deep CNN.

Given the complexity and richness of many auditory stimuli, particularly music, generating spectrograms of reasonable dimensions for a CNN necessarily involves an important compromise between frequency resolution and the length of the signal vector that can be encoded. In order to explore this compromise, we also generate logSTFT spectrograms at a lower sampling rate (2kHz vs 8 kHz).

The presence of clear visual texture representing audio features can be seen in the spectral and randomly transformed representations of a variety of sounds (Figure 1). Input representations obtained via RMT are significantly different from the highly structured spectral representations obtained via Fourier transform. One notable feature is that the information density is significantly higher in the case of RMTs.

### 2.3 Description of Network

We use a deep convolutional network design similar to those used in imageNet challenges [4]. Training was performed on NVidia Quadro M2000M GPU, which constrained the number of convolutional layers. Zero-mean variance-normalized data in mini batches of 50 was loaded to the input layer of size 204x204. Each is processed with 6 convolutional layers of size 3x3, each followed by subsequent max pooling layer of size 2x2. The output of the convolutional layers was passed to 3 fully connected layers with 30% dropout. Classification was obtained by application of the soft max function to the final output. Rectified nonlinearity and Xavier initialization were used for both convolutional and fully connected layers. The neural net was implemented in Python 3.5 using Theano and Lasagne libraries.

Table 1 – Structure of CNN

| #  | Output size | Layer type      | Filter size |
|----|-------------|-----------------|-------------|
| 1  | 1x204x204   | Input           |             |
| 2  | 32x202x202  | Convolutional   | 3x3         |
| 3  | 32x101x101  | Max pooling     | 2x2         |
| 4  | 32x99x99    | Convolutional   | 3x3         |
| 5  | 32x50x50    | Max pooling     | 2x2         |
| 6  | 64x48x48    | Convolutional   | 3x3         |
| 7  | 64x24x24    | Max pooling     | 2x2         |
| 8  | 64x22x22    | Convolutional   | 3x3         |
| 9  | 64x11x11    | Max pooling     | 2x2         |
| 10 | 128x9x9     | Convolutional   | 3x3         |
| 11 | 128x5x5     | Max pooling     | 2x2         |
| 12 | 128x3x3     | Convolutional   | 3x3         |
| 13 | 128x2x2     | Max pooling     | 2x2         |
| 14 | 512         | Fully connected |             |
| 15 | 256         | Fully connected |             |
| 16 | 75          | Fully connected |             |
| 17 | 75          | Soft max        |             |
| 18 | 1           | Output          |             |

## 3 Experiments

### 3.1 Supervised feature learning from custom audio dataset

2D inputs were generated according to each of the above-described transforms from audio files that contain > 5 compositions per composer. There is significant redundancy in the datasets as the overlap between neighboring spectrograms is 80% (20% shift of the window). These files were sampled randomly to generate 1000 spectrograms for each class. The result was saved as a set of the complex number arrays. The dataset was shuffled and divided into three parts, consisting of 60%, 20% and 20% to be referred to as training, cross validation and testing sets respectively. Training was performed on the training dataset until convergence, which took 150 epochs, of roughly 1.5 minutes each. Given the ease of additional labeled data generation, unlike image data, it was not necessary to use jitter to expand the dataset. Additional spectrograms, sharing similar texture patterns, could have been sampled from the same audio files if needed.

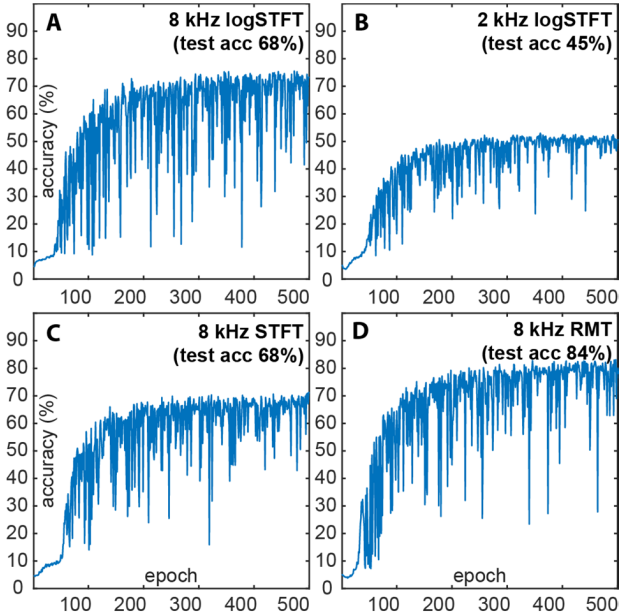

**Figure 2** – Error on cross validation dataset as a function of training epoch of music obtained via (A) logSTFT (8kHz dataset), (B) logSTFT (2kHz dataset) (C) STFT and (D) RMT both (8kHz dataset)

### 3.2 Classification of the full dataset

Once trained, we applied the CNN to the testing datasets, achieving a classification accuracy of around 68% for both the logSTFT dataset at 8kHz (vs. 48% at 2kHz) and the linear STFT dataset after ~150 epochs (Figure 2 A,B and C). Performance was highest for the dataset obtained via RMT (80% - Figure 2 D). One may argue that this classification accuracy is relatively low, however it should be kept in mind that 5 s chunks of audio files may not be representative of musical style (i.e. they lie at the beginning or the end of the file, pauses, low information sections, etc.). Indeed we expect that this task would be challenging even for human classifiers, but the human performance baseline has not yet been tested. To make sense of the classification errors, we plotted the confusion matrix (Figure 3).

## 4 Relation to other work

Learning of audio features and classification from spectral representations of sound via deep convolutional networks is an expanding field of research. To date, a number of papers have approached the problem of Acoustic Scene Classification (ASC), with only a handful studying musical features with convolutional neural networks [9]. In one study with an optimized DCNN, the authors report an accuracy score of 0.69 in the classification of the DCASE 2013 database (containing limited training samples) using an alternative spectral representation and with significantly shorter 1-s clips from audio files. In another study, deep belief networks were successfully used for unsupervised learning of audio features from a limited dataset of very short audio clips. [10]. The closest work that we are aware of is the excellent work by Sprengel et al. on the BirdCLEF database, in which only spectrograms obtained via STFT (filter distribution unspecified) are used to classify bird species by calls with ~69% accuracy [11]. Unfortunately, the MIREX dataset of music from 11 classical composers is not publicly available, however the highest performance recorded on this smaller dataset was in the range of 75-80%.

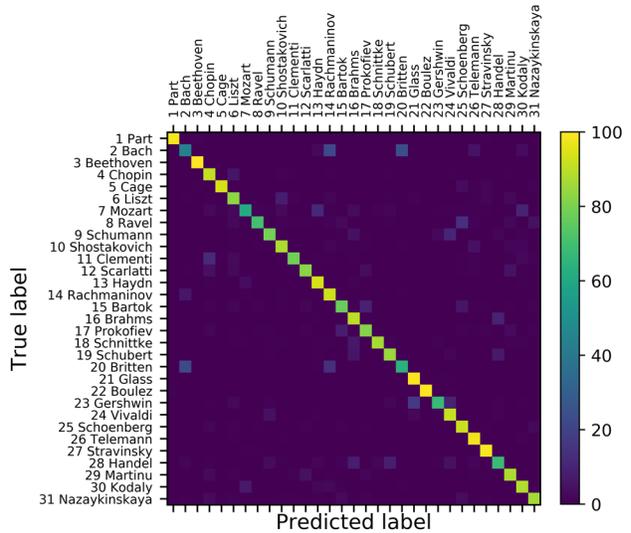

**Figure 3** – Confusion Matrix for the RMT derived dataset

## 5  Conclusions

In this paper we have further demonstrated the value of CNNs in the classification of audio data. We were able to apply techniques developed over the last five years in the image-processing field to a large and relatively novel type of dataset, achieving very high classification performance. It would be interesting to compare the algorithm's performance on this task with that of humans. Given the length of the audio clips used, we expect that this task would be very difficult for even musically trained humans. Interestingly, the highly structured spectrograms obtained via STFT were inferior, as a basis for classification, to the representation obtained by the random matrix transform of raw waveforms.